\documentclass[preprint,showpacs,aps,prd,nofootinbib,letterpaper]{revtex4-1}
\usepackage{graphicx}
\usepackage{subfigure}
\usepackage{dcolumn}
\usepackage{amsmath}
\usepackage{amssymb}
\usepackage{times}
\usepackage{epsfig}
\usepackage{float}     
\usepackage{relsize}
\usepackage{mathrsfs}
\RequirePackage{xspace}
\begin{document}
\title{ Exclusive $J/\psi$ Production in Diffractive Process with AdS/QCD Holographic Wave Function in BLFQ }
\author{Ya-ping Xie}\email{xieyaping@impcas.ac.cn}
\affiliation{Institute of Modern Physics, Chinese Academy of
Sciences, Lanzhou 730000, China}
\affiliation{Department of Physics, Lanzhou University, Lanzhou 730000, Chnia}

\author{Xurong Chen}\email{xchen@impcas.ac.cn}
\affiliation{Institute of Modern Physics, Chinese Academy of
Sciences, Lanzhou 730000, China}
\begin{abstract}
The AdS/QCD holographic wave function of basis light-front quantization (BLFQ) for vector meson $J/\psi$ is applied in this manuscript. The exclusive production of $J/\psi$ in diffractive process is computed in dipole model with AdS/QCD holographic wave function. We use IP-Sat and IIM model in the calculation of the differential cross section of the dipole scattering off the proton. The prediction of AdS/QCD holographic wave function in BLFQ gives a good agreement to the experimental data.
\end{abstract}
 %1 4.80.Bn
\maketitle
\section{Introduction}
\indent Anti-de Sitter (AdS) and quantum chromodynamics (QCD) has been applied successfully in various fields~\cite{Maldacena:1997re,Karch:2006pv,Erdmenger:2007cm,Forshaw:2012im,Costa:2013uia}. In this manuscript, we use the AdS/QCD holographic wave function to calculate the $J/\Psi$ production in photonproduction process.\\
\indent The exclusive vector meson production at the HERA are a good probe for the structure of hadrons~\cite{Ivanov:2004ax}. There are various approaches to compute the vector meson production in diffraction~\cite{Ryskin:1992ui,Brodsky:1994kf,Collins:1996fb}. The dipole model was applied successfully to calculate the production of the vector mesons in diffractive process~\cite{Nikolaev:1990ja,Mueller:1993rr}. According to the dipole picture, the virtual photon fluctuates into quark and antiquark pair which is called dipole in the diffractive process firstly. Then, the dipole scatters off the proton by exchange gluons. Finally, the dipole recombines a vector meson. Thus, the amplitude of the vector meson production in the diffractive process contains three parts, they are the light-cone wave function of the photon, the cross section of the dipole scattering off the proton and wave function of the vector meson. The light-cone wave function of the photon can be calculated in QED, and the differential cross section of the dipole scattering off the proton was firstly proposed and fitted by the experimental data several years ago~\cite{GolecBiernat:1998js,GolecBiernat:1999qd}.\\
 \indent The wave function of the vector meson is very important in the photonproduction in the diffractive process. It is a non-perturbative problem. The wave function of vector meson can not be computed analytically. Thus, it is modeled after the photon wave function. There are various models for the wave function of the vector mesons, for example, NNPZ, DGKP, Boosted Gaussian and Gaus-LC~\cite{Dosch:1996ss,Forshaw:2003ki,Goncalves:2004bp,Nemchik:1994fp,Nemchik:1996cw,Kowalski:2006hc}. The wave function of vector meson has free parameters, which are determined by the decay constant and normalization condition. But, they can not give the spectrum of the heavy quarkonium states. The wave functions of heavy quarkonium was studied in non-relativistic potential model by Cornell group in 1970s~\cite{Eichten:1978tg,Eichten:1979ms}. The Cornell potential model gives a good description to the spectrum of the charmonium states. But it can not reproduce the decay width of the charmonium states. On the other side, the wave function of heavy quarkonium were studied in decretized momentum basis~\cite{Brodsky:1997de} and basis of light-front quantization (BLFQ)
~\cite{Vary:2009gt,Li:2015zda,Wiecki:2015,Honkanen:2010rc}. The AdS/QCD holographic wave function in BLFQ can reproduce the decay width of the charmonium states, and it is in the light-cone system. Thus, we think it is also valid in the photonproduction in diffraction. We should reproduce the $J/\psi$ production in $\gamma^*p\to J/\psi p$.\\
\indent In this manuscript, we use the AdS/QCD holographic wave function in BLFQ and dipole model to compute the prediction of the cross section of the $J/\psi$ in diffractive process, and compare the result with the experimental data.\\
\indent The differential cross section of the dipole scattering off the proton is also included in the amplitude in the diffractive process. In the literature, the IP-Sat and IIM models are successfully to describe the process of the differential cross section of the dipole scattering off the proton~\cite{Kowalski:2003hm,Bartels:2002cj,Kowalski:2006hc,Iancu:2003ge,Soyez:2007kg,Watt:2007nr,Rezaeian:2012ji,Rezaeian:2013tka}. The two models both have free parameters which are determined from the fit to HERA experimental data. There are various parameter sets for IP-Sat model and IIM model.\\
 \indent In this manuscript, we apply the AdS/QCD holographic wave function in BLFQ of the $J/\psi$ in the diffractive process. Then, we reproduce the cross section of the $J/\psi$, and compare our prediction with the experimental data of HERA. This paper is organized as follow. The brief review of the dipole model and AdS/QCD holographic wave function in BLFQ will be presented in Sec.~II. The numerical results and some discussion will be presented in Sec.~III, and the conclusion will be presented in Sec.~IV.
\section{Review of dipole cross section and AdS/QCD wave function}
\subsection{IP-Sat model and IIM model}
We begin with formulas of the differential cross section of the $\gamma^*p\to Vp$ in the diffractive process. The total cross section can be computed by integrating $t$. The differential cross section of vector meson in diffractive process can be computed as following:
\begin{eqnarray}
  \frac{d\sigma^{\gamma^* p\to Vp}}{dt}
 =\frac{R_g^2(1+\beta^2)}{16\pi}|\mathcal{A}_{T,L}(x_p,Q^2,\Delta)|^2,
\label{dsdt}
\end{eqnarray}

where $x_p=(Q^2+M_V^2)/(Q^2+W^2)$ in the diffractive process, and $t=-\Delta^2$, $T,L $ denote the transverse and longitudinal amplitude. The $1+\beta^2$ is the real part of the amplitude,
\begin{equation}
\beta=\tan(\frac{\pi}{2}\lambda),
\end{equation}
where $\lambda$ is calculated as
\begin{equation}
\lambda=\frac{\partial \ln (\mathrm{Im}\mathcal{A}_{T,L}(x))}{\partial
\ln1/x}.
\end{equation}
The factor $R_g^2$ is the about the skewness effect~\cite{Shuvaev:1999ce}, it gives
\begin{equation}
R_g=\frac{2^{2\lambda+3}}{\sqrt{\pi}}\frac{\Gamma(\lambda+5/2)}{\Gamma(\lambda+4)}.
\end{equation}
The amplitude $\mathcal{A}^{\gamma^*p\to Vp}_{T,L}(x_p, Q^2, \Delta)$ in Eq.~(\ref{dsdt}) is

\begin{eqnarray}
\mathcal{A}_{T,L}^{\gamma^*p\to Vp}(x_p, Q^2,\Delta)= i\int
d^2r\int_0^1\frac{dz}{4\pi} \int
d^2b(\psi_V^*\psi_{\gamma})_{T,L}(z,r,Q^2)e^{-i(b-(1-z)r)\cdot
\Delta }\frac{d\sigma_{q\bar{q}}}{d^2b},
\end{eqnarray}
where $z$ is the fraction of the momentum carried by quark to the photon, $(\psi^*_V\psi_\gamma)_{T,L}(z,r,Q^2)$ is the overlap of the photon and the vector meson, and $\frac{d\sigma_{q\bar{q}}}{d^2b}$ is the differential dipole cross section.
In the IP-Sat model, the differential dipole cross section is\cite{Kowalski:2006hc,Rezaeian:2012ji}
 \begin{equation}
 \frac{d\sigma_{q\bar{q}}}{d^2b}=2[1-\exp(-\frac{1}{2\pi B_p}\frac{\pi^2}{2N_c}r^2\alpha_s(\mu^2)xg(x, \mu^2)T_p(b)],
 \end{equation}
 where $T_p(b)=\exp(-b^2/2B_p)$ is the profile function, and $xg(x,\mu^2)$ is the gluon density, which is evolved from $\mu_0^2$ to $\mu^2=\mu_0^2+4/r^2$ by leading order DGLAP equation, the initial condition of the gluon density is 
 \begin{eqnarray}
 xg(x,\mu_0^2)=A_gx^{-\lambda_g}(1-x)^{5.6}.
 \end{eqnarray} 
In the IIM model, the differential dipole cross section is written as \cite{Iancu:2003ge,Watt:2007nr,Rezaeian:2013tka}
 \begin{eqnarray}
 \frac{d\sigma_{q\bar{q}}}{d^2b}=2\times\begin{cases}
 \mathcal{N}_0(\frac{rQs}{2})^{2(\gamma_s+(1/\kappa\lambda Y)\ln(2/rQs))},\quad\! rQs\le2,\\
 1-\exp\big(-a\ln^2(b rQs)\big),\quad\quad rQs>2,
 \end{cases}
 \end{eqnarray}
 where $Qs=(x/x_0)^{\lambda/2}\exp(-\frac{b^2}{4\gamma_sB_p})$, $\kappa=9.9$, and $Y=\ln(1/x)$. The parameters of the IP-Sat model and IIM model are determined from the fit to combined HERA data for the reduced cross section, the parameters we used are the same as Ref.~\cite{Rezaeian:2012ji, Rezaeian:2013tka}.\\
 \subsection{AdS/QCD holographic wave function in BLFQ}
\indent The $\psi^*_V\psi_\gamma$ is the overlap of the photon and vector meson, we use the AdS/QCD wave function in the basis of light-front quantization. The light vector meson is not considered in this manuscript, because this approach is not applicable for the light vector meson. For more detail information about the AdS/QCD holographic wave function in BLFQ we refer the readers to the Ref.~\cite{Li:2015zda}. The wave function of the heavy quarkonium is eigenfunction of the eigenvalue equation $H_{eff}\psi^J_{m_j}=M^2_V\psi^J_{m_j}$, where $J$, $m_j$ are total spin and magnetic spin.
The Hamiltonian is 
\begin{equation}
H_{eff}=q_\perp^2+\kappa^4\zeta_\perp^4+\frac{m_q^2}{z}+\frac{m^2_{\bar{q}}}{(1-z)}
-\frac{\kappa^4}{(m_q+m_{\bar{q}})^2}\partial_z(z(1-z)\partial_z)+V_g,
\end{equation}
where $z$ is the fraction of the momentum carried by the quark $z=p_q^+/P^+ $. The variable $\kappa$ is the strength of the potential , and $m_q$ is the effective quark mass, they are free parameters to fit from the spectrum of the charmonium states. The relative transverse momentum is $k_\perp=p_{q\perp}-zP_\perp$, and $\zeta_\perp$ is the holographic coordinate $\zeta_\perp=\sqrt{z(1-z)}r_\perp,$ with $r_\perp$ the radius of the two quarks. The conjugate of $\zeta_\perp$ is defined as $q_\perp=k_\perp/\sqrt{z(1-z)}$. $V_g$ is the one gluon exchange potential between the quark and antiquark, the detail information of $V_g$ can be referred to the Ref.~\cite{Li:2015zda}.
The AdS/QCD holographic wave function in BLFQ of charmonium states can be written as 
\begin{equation}
\psi^J_{m_j}(z,\mathbf{k}_\perp,s,\bar{s})=\sum_{m_j,m+s+\bar{s}}
\widetilde{\psi}_{m_j}^J(n,m,l,s,\bar{s})\phi_{nml}(\mathbf{k}_\perp/\sqrt{z(1-z)},z).
\end{equation}
where $\widetilde{\psi}_{m_j}^J(n,m,l,s,\bar{s})$ is the eigenfunction of the hamiltonian equation, and $s, \bar{s}$ are the helicities of the quark and antiquark, $n$ is radial quantum number, $m$ is the angular momentum quantum number. The $\phi_{nml}(\mathbf{k}_\perp/\sqrt{z(1-z)},z)$ is the product of 2D harmonic basis function $\phi_{nml}(q_\perp)$ and $\chi_l(z)$.
They are 
\begin{equation}
\phi_{nm}(q_\perp)=\frac{1}{\kappa}\sqrt{\frac{4\pi n!}{(n+|m|!)}}(q_\perp/\kappa)^{|m|}
e^{-\frac{1}{2}q_\perp^2/\kappa^2}
L_n^{|m|}(q_\perp^2/\kappa^2)e^{im\theta},
\end{equation}
where $L_n^m(x)$ is the Laguerre polynomial.
The $\chi_l(z)$ is
\begin{eqnarray}
\chi_l(z)=\sqrt{4\pi(2l+2\mu+1)}\frac{\sqrt{l!
\Gamma(l+2\mu+1)}}{\Gamma(l+\mu+1)}z^{\mu/2}(1-z)^{\mu/2}P_l^{(\mu,\mu)}(2z-1),
\end{eqnarray}
where $\mu=4m_q^2/\kappa^2$, and the $P_l^{a,b}(x)$ is the Jaccobi polynomal. The fourier transformation of 2D harmonic basis function is defined as
\begin{eqnarray}
\widetilde{\phi}_{nnml}(r_\perp)&=&\int\frac{d^2q_\perp}{2\pi^2}e^{-iq_\perp\cdot r_\perp}
\phi_{nml}(q_\perp)\notag\\&=&
\kappa\sqrt{\frac{ n!}{\pi(n+|m|!)}}(\kappa r_\perp)^{|m|}
e^{-\frac{1}{2}\kappa^2r_\perp^2}
L_n^{|m|}(\kappa^2r_\perp^2)e^{im\theta}(-1)^ni^{|m|}.
\end{eqnarray}
Finally, the AdS/QCD holographic wave function in BLFQ in coordinate space can be written as
\begin{eqnarray}
\widetilde{\psi}^V_{s,\bar{s}}(z,r_\perp)=\sqrt{z(1-z)}\sum_{n,m,l}\widetilde{\psi}^J_{m_j}(n,m,l,s,\bar{s})\widetilde{\phi}_{nml}(\sqrt{z(1-z)r_\perp})\chi_l(z).
\end{eqnarray}
\indent The light-cone wave function of the virtual photon can be computed from QED, the longitudinal virtual photon ($m=0$) is given by~\cite{Forshaw:2003ki}

\begin{equation}
  \psi_{s,\bar{s},m=0}(r,z,Q)=e_fe\sqrt{N_c}\delta_{s,-\bar{s}}2Qz(1-z)\frac{K_0(
  \epsilon r)}{2\pi}.
\end{equation}
The transverse virtual photon ($m=\pm1$) reads
\begin{eqnarray}
  \psi_{s,\bar{s},m=\pm1}(r,z,Q)=\mp e_fe\sqrt{2N_c}\{ie^{\pm
  i\theta}[z\delta_{s,\pm}\delta_{\bar{s},\mp}-(1-z)\delta_{s,\mp}
  \delta_{\bar{s},{\pm}}]\partial_r \mp
  m_q\delta_{s,\pm}\delta_{\bar{s},\pm}\}\frac{K_0(\epsilon r)}{2\pi},\notag\\
\end{eqnarray}
where $\epsilon=\sqrt{z(1-z)Q^2+m_q^2}$, $N_c$ is the color number, and $K_0(x)$ and $K_1(x)$ are second kind Bessel functions.
If the eigenvalue equation is solved and the numerical expression of the $\widetilde{\psi}^J_{m_j}(n,m,l,s,\bar{s})$ is providing, we can compute the overlap of the $\psi^*_V\psi_\gamma(r,z)_{T,L}$. Then, we can calculate the differential cross section of the diffractive process.\\
\indent The Boosted Gaussian model is a successful ansatz~\cite{Forshaw:2003ki,Kowalski:2006hc}. It is modeled after the structure of the photon, its transversely polarized vector meson function is
\begin{equation}
\psi^V_{s,\bar{s}, m=\pm1}=\mp\sqrt{2N_c}\frac{1}{z(1-z)}\{ie^{\pm i\theta}[z\delta_{s,\pm}
\delta_{\bar{s},\mp}-(1-z)\delta_{s},\mp\delta_{\bar{s},\pm}]\partial_r\mp m_q\delta_{s,\pm}
\delta_{\bar{s},\pm}\}\phi_T(z,r).
\end{equation}
The longitudinal polarized wave function is some different from the photon. It is given by
\begin{eqnarray}
\psi^V_{s,\bar{s}, m=0}=\sqrt{N_c}\delta_{s,-\bar{s}}\Big[M_V+
\frac{m_q^2-\nabla_r^2}{M_Vz(1-z)}\Big]\phi_L(z,r),
\end{eqnarray}
where $M_V$ is the mass of the vector meson, and $\nabla_r^2=(1/r)\partial_r+\partial^2_r$. The transversely and longitudinally scalar function of the Boosted Gaussian take the same form, it is written 
\begin{eqnarray}
\phi^V_{T,L}(z,r)=\mathcal{N}_{T,L}z(1-z)\exp\Big(-\frac{m_q^2\mathcal{R}^2}{8z(1-z)}
-\frac{2z(1-z)r^2}{\mathcal{R}^2}+\frac{m_q^2\mathcal{R}^2}{2}\Big),
\end{eqnarray} 
where $\mathcal{N}$ and $\mathcal{R}$ are free parameters to be determined from the normalization and decay width conditions.\\
\indent We can see that there are many advantages for the AdS/QCD holographic wave function in BLFQ. The Boosted Gaussian model is just a asantz, the AdS/QCD holographic wave function in BLFQ is from the first principle. The Boosted Gaussian model can not reproduce the spectrum of the charmonium states, the AdS/QCD holographic wave function in BLFQ can reproduce the spectrum. It is necessary to introduce other parameter for the excited states for Boosted Gaussian model. But the parameters are same for the excited states in the AdS/QCD holographic wave function in BLFQ.
\section{Numerical result and discussion}
\begin{figure}[H]
\centering
\includegraphics[width=3in]{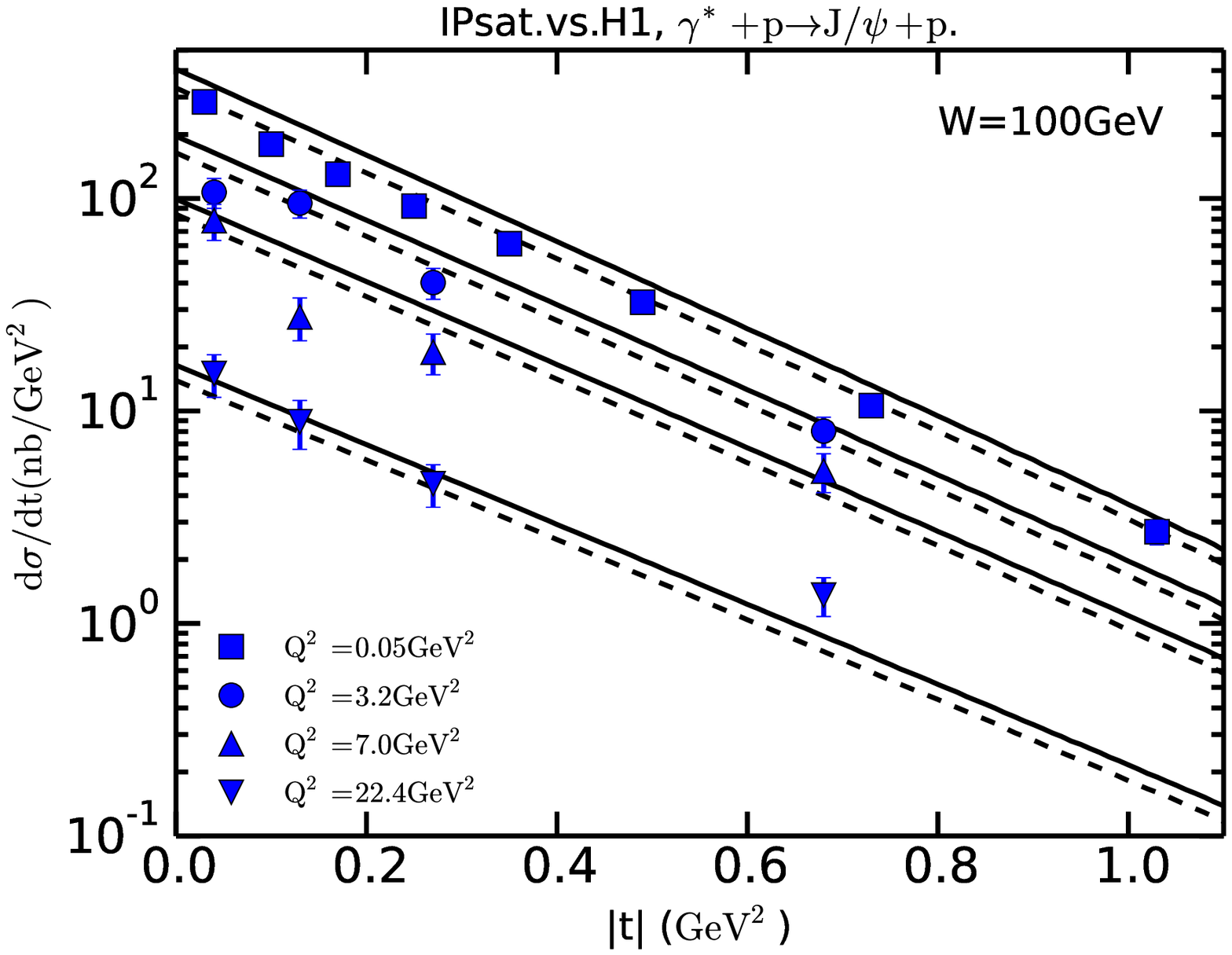}
\includegraphics[width=3in]{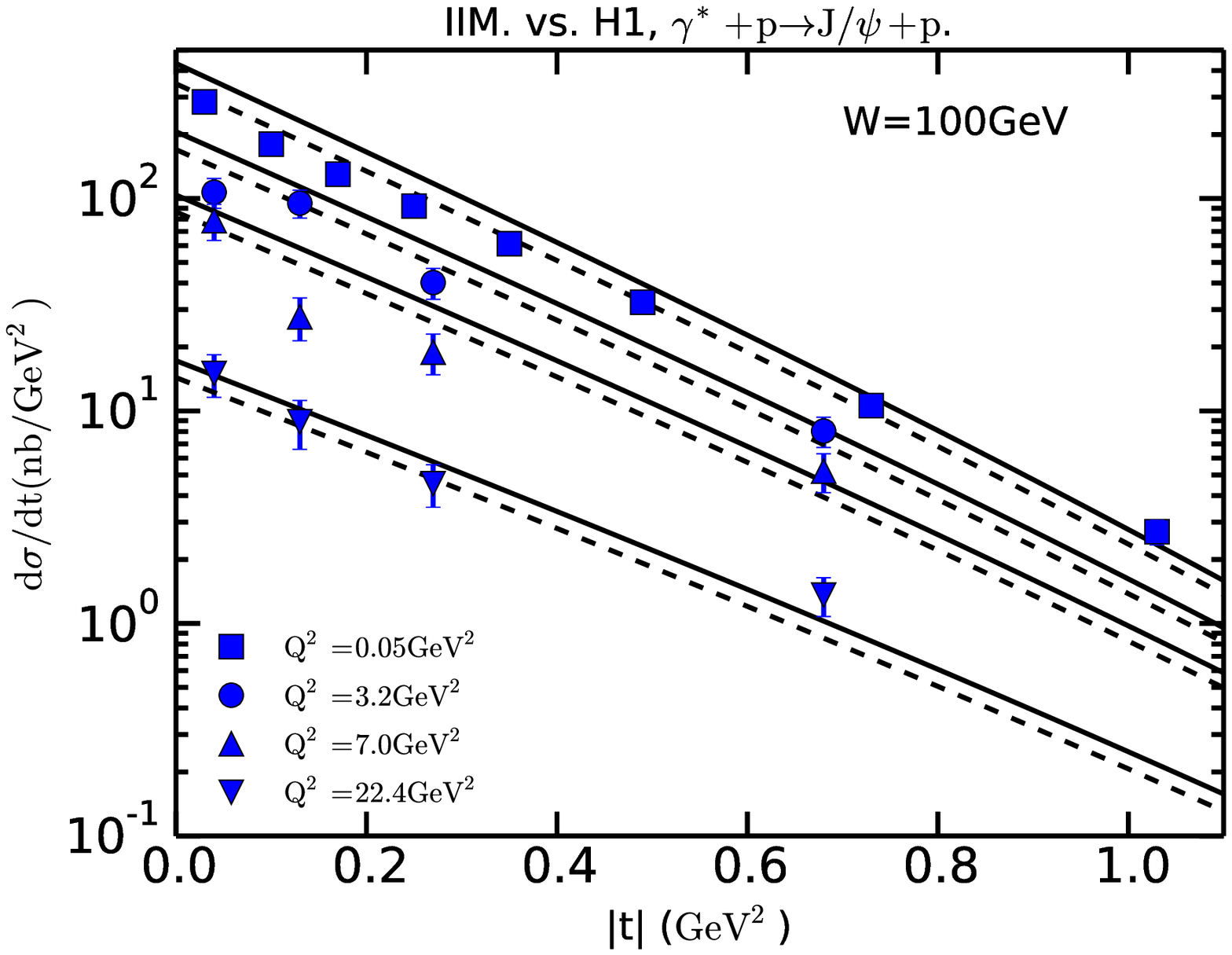}
\caption{(Color online) The differential cross section of $J/\psi$ in diffractive process using IP-Sat and IIM model as a function of $|t|$. The solid curve is using the parameter set with $m_c=$1.27~GeV with Boosted Gaussian model. The dashed curves are using the parameter set with $m_c$=1.27~GeV with AdS/QCD holographic wave function in BLFQ. The experimental data are taken from Refs.~\cite{Aktas:2005xu}.}\label{fig2}
\centering
\end{figure}

\begin{figure}[htbp]
\centering
\includegraphics[width=3in]{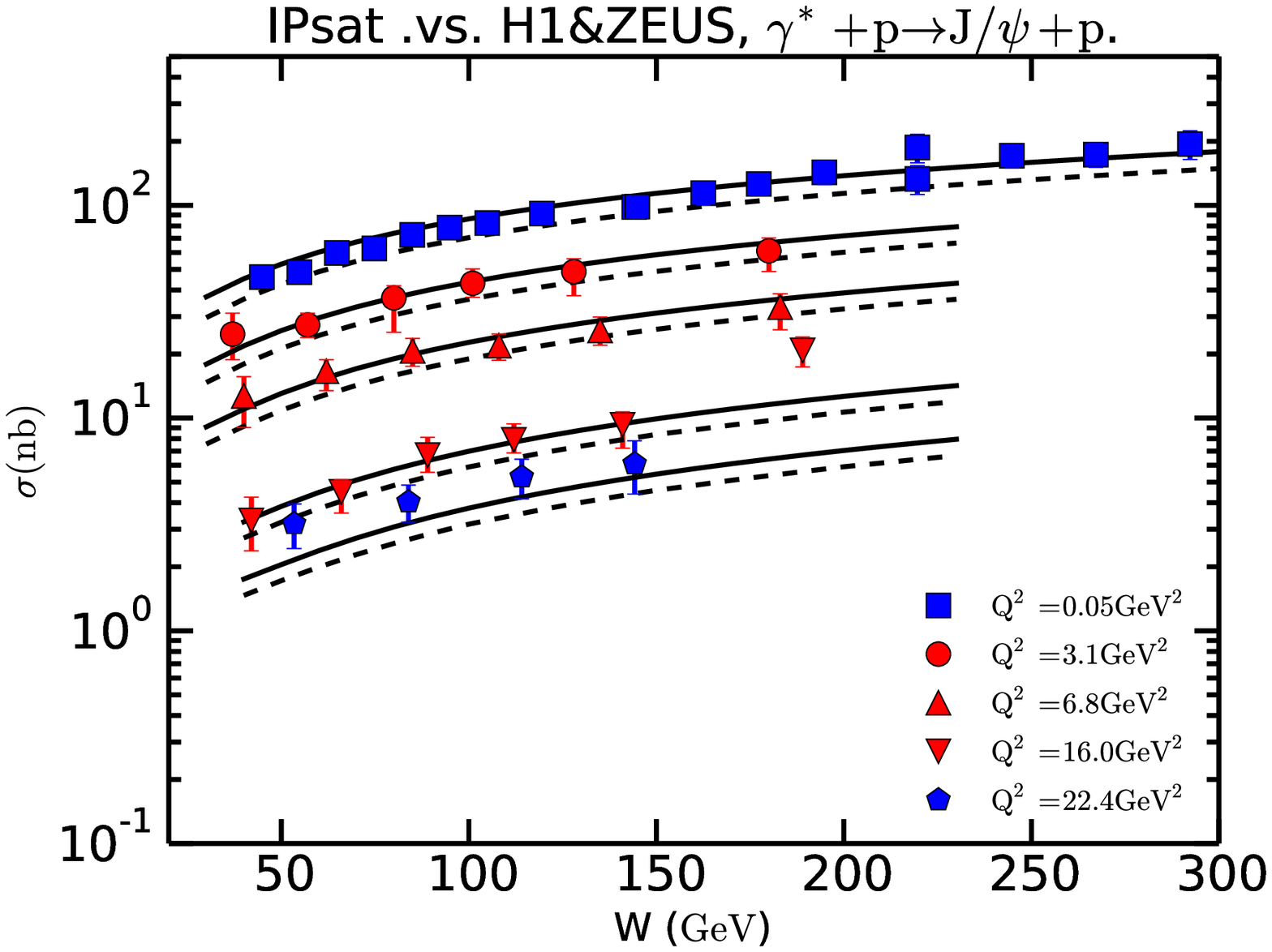}
\includegraphics[width=3in]{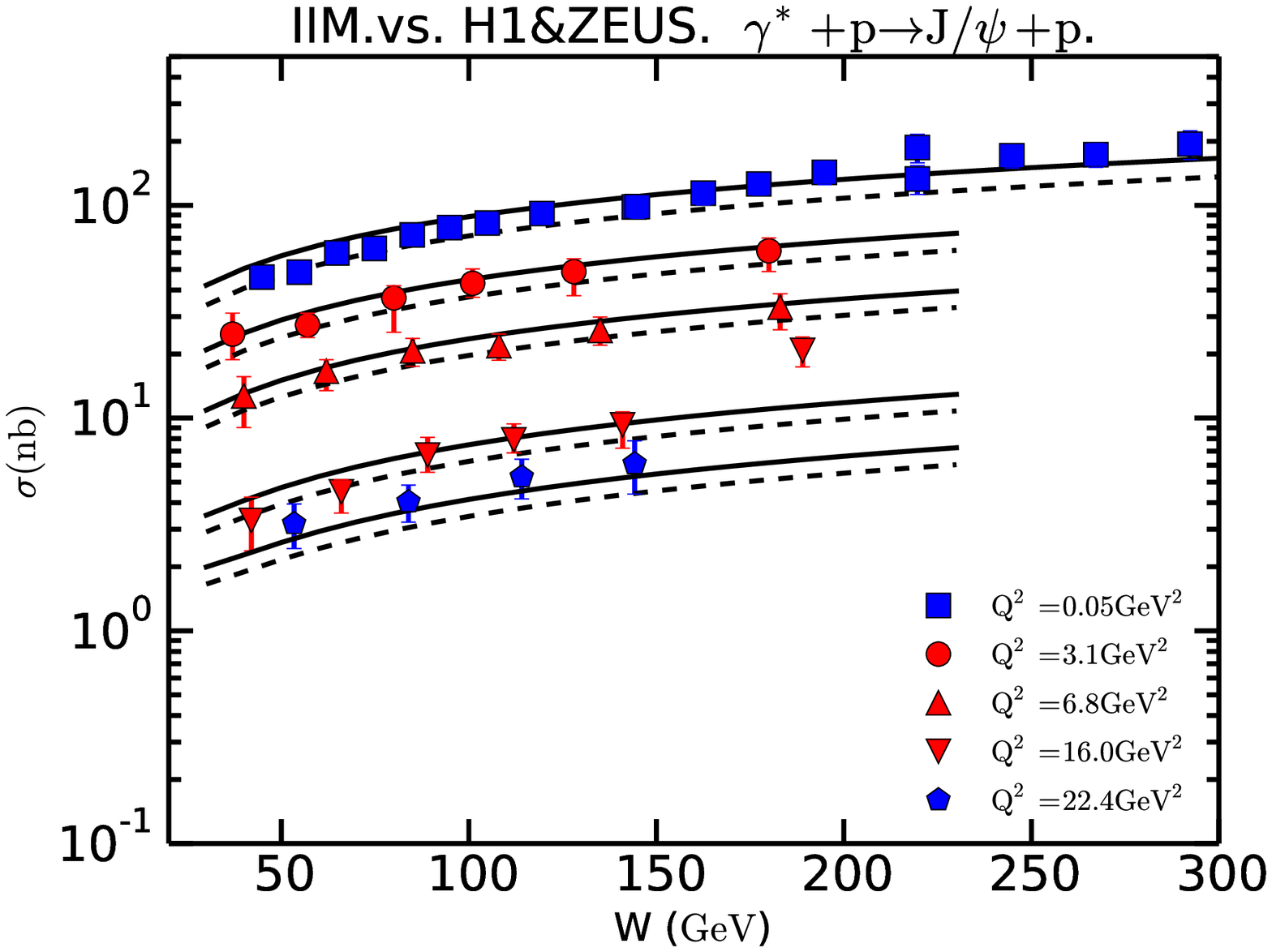}
\caption{(Color online) The total $J/\psi$ cross section as a function of W compared with H1 and ZEUS experimental data~\cite{Aktas:2005xu,Chekanov:2004mw} from IP-Sat and IIM model. The solid curve is using the parameter set with $m_c=$1.27~GeV with Boosted Gaussian model. The dashed curves are using the parameter set with $m_c$=1.27~GeV with AdS/QCD holographic wave function in BLFQ. }
\label{fig3}
\end{figure}

\begin{figure}[htbp]
\centering
\includegraphics[width=3in]{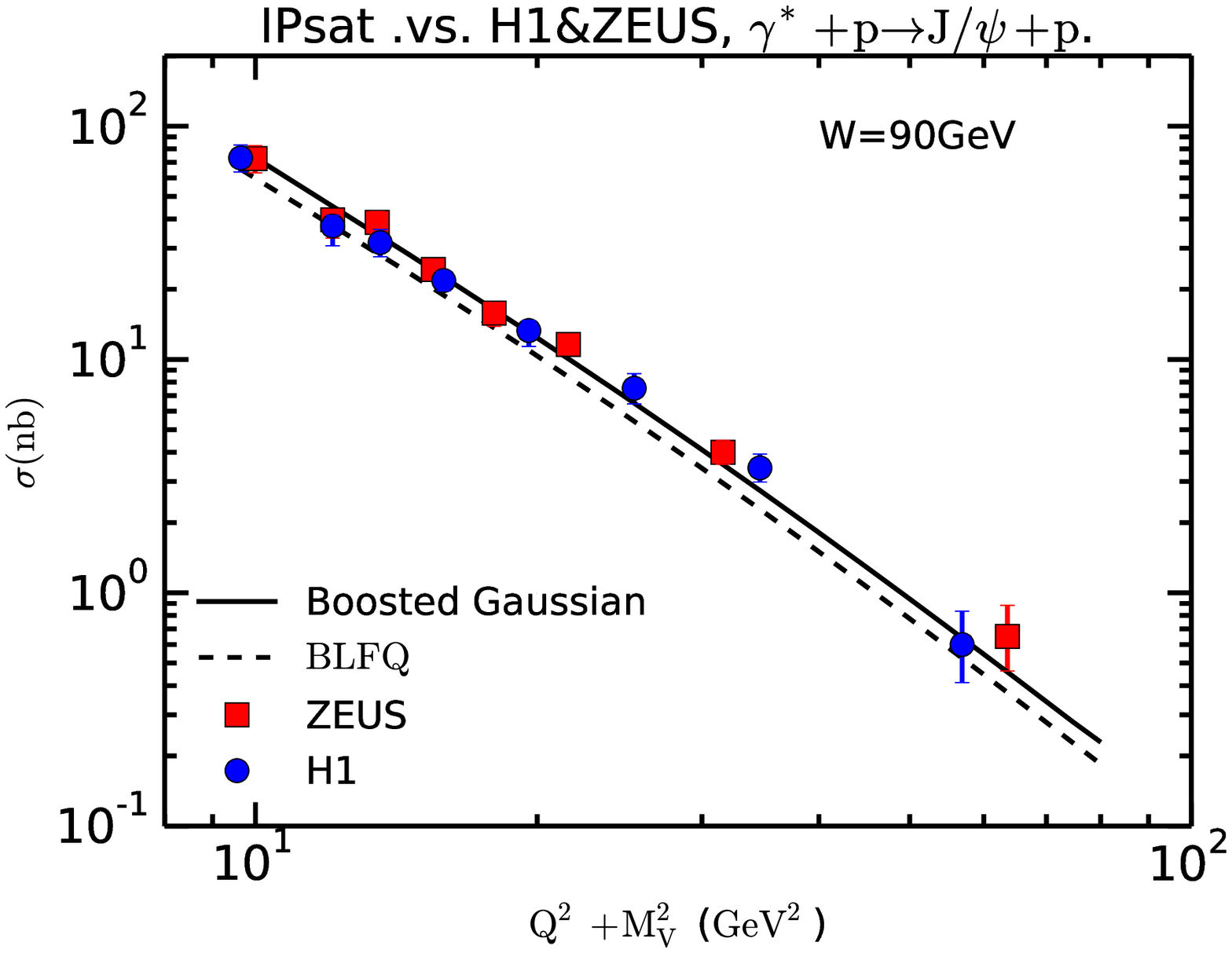}
\includegraphics[width=3in]{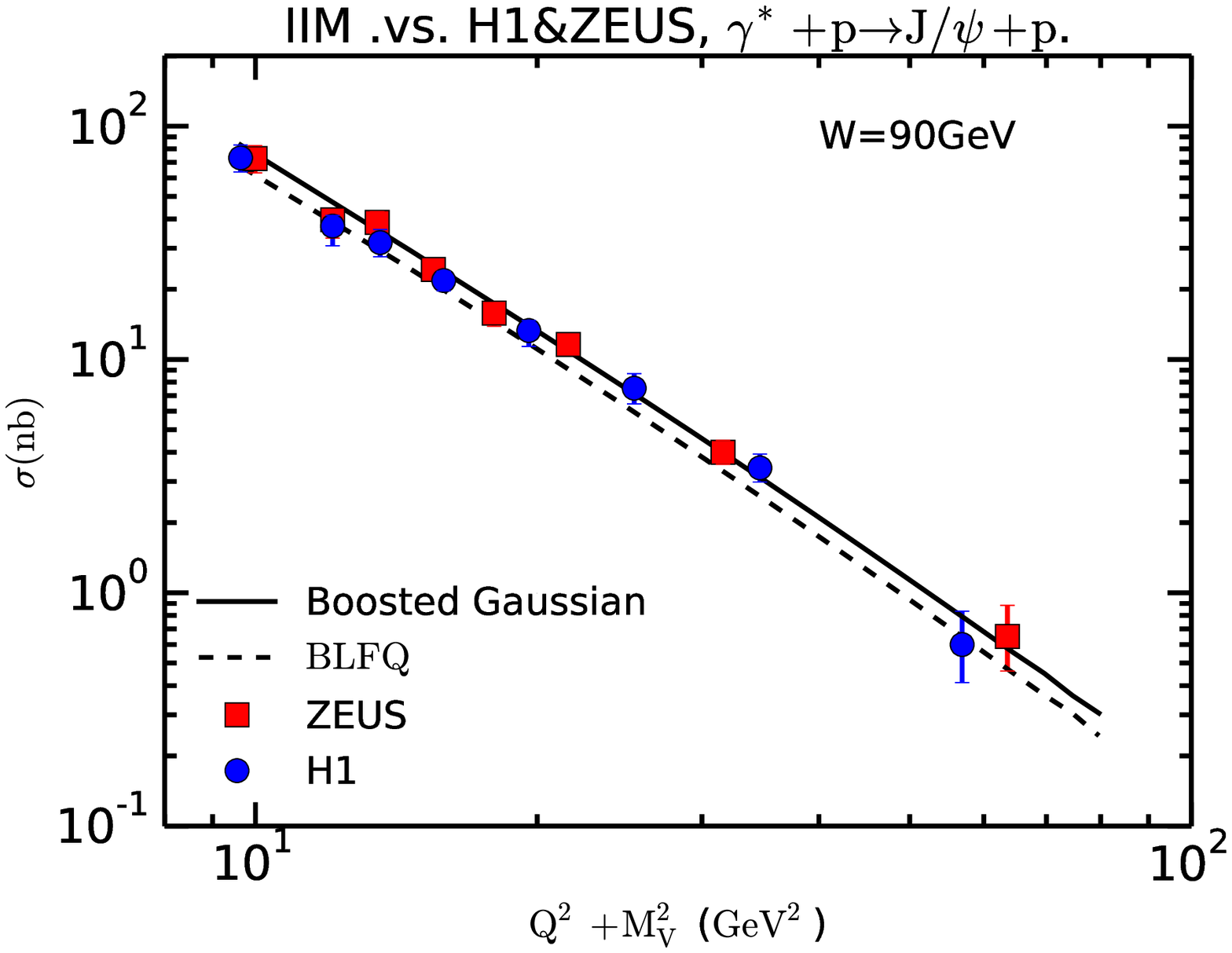}
\caption{(Color online) (Color online) The total $J/\psi$ cross section as a function of $Q^2+M_V^2$ compared with H1 and ZEUS experimental data~\cite{Aktas:2005xu,Chekanov:2004mw} from IP-Sat model and IIM model. The solid curve is using the parameter set with $m_c=$1.27~GeV with Boosted Gaussian model. The dashed curves are using the parameter set with $m_c$=1.27~GeV with AdS/QCD holographic wave function in BLFQ.}
\label{fig4}
\end{figure}

We calculate the differential cross section and total cross section using IP-Sat and IIM model with the AdS/QCD holographic wave function in BLFQ with parameters with Nmax=8, $m_q=1.492$~GeV, and $\kappa=0.963$~GeV, where the quark mass in the holographic wave function in BLFQ is different from the quark mass of dipole and the photon wave function. The differential cross section are showed in Fig.~\ref{fig2}. The total cross section are presented in Fig.~\ref{fig3} and Fig.~\ref{fig4}. We integrate the $|t|$ up to 1~$\mathrm{GeV}^2$ with $m_c=1.27$~GeV parameter set for IP-Sat and IIM model when we calculate the total cross section. \\
\indent The differential cross section of the $J/\psi$ in the diffractive process are presented in Fig.~\ref{fig2},
the differential cross section are computed in two wave functions and compared with the experimental data of H1. The blue markers are H1 data. The solid curves are computing using Boosted Gaussian wave function whose parameters can be found in Ref.~\cite{Armesto:2014sma}. The dashed curves are computing using AdS/QCD holographic wave function in BLFQ. It is seen that the two wave functions both give a good agreement to the experimental data.\\
The total $J/\psi$ cross section in diffractive process are showed in Fig.~\ref{fig3} and Fig.~\ref{fig4}. IP-Sat and IIM model both are applied in the calculation. The W dependence cross section at a fixed $Q^2$ are presented in Fig.~\ref{fig3}. It is easy to find that the results of parameter set with $m_c=$1.27~GeV give a good description of the experimental data. The cross section as a function of $Q^2$ at a fixed W are showed in Fig.~\ref{fig4}. IP-Sat and IIM model are both applied in the calculation. The two wave functions both give a good agreement to the experimental data of HI and ZEUS.\\
\indent At the end of the day, we can see that the AdS/QCD holographic wave function in BLFQ gives a good description to the differential and total cross section of $J/\psi$ in diffractive process. The holographic wave function gives a little lower prediction than the Boosted Gaussian wave function with same dipole model parameters.
\section{conclusion}
\indent In this manuscript, we use AdS/QCD holographic wave function in calculation of $J/\psi$ production in diffractive process. We compute the prediction of differential and total cross section of $J/\psi$ using IP-Sat and IIM model with AdS/QCD holographic wave function in BLFQ and compare the results with the experimental data. There are two parameter sets in IP-Sat and IIM model, the values of mass of the charm quark are different in the two parameter sets. The results show that the prediction using the parameter  with $m_c=$1.27~GeV give a good description to the experimental data in the small $Q^2$. The two wave functions both give a good prediction in $\gamma^*p\to J/\psi p$. In the work, we only consider the production of ground state of charmonium. The excited states of charmonium such as $\psi(2s)$, $\psi(3s)$ are absent in the calculation. The parameters of the excited states are the same as the ground state, it is not necessary to introduce new parameters for the excited states, the excited states of the heavy vector meson will be considered at the next step.
\section{Acknowledgements}
We thank the authors of Ref.~\cite{Li:2015zda} for providing the data file of the wave function, and thank
 communication with H.~Mantysaari and A.~H.~Rezaenian. This work is supported in part by the National 973 project in China (No:~2014CB845406).

\end{document}